\documentstyle[prl,aps,multicol,epsf]{revtex}

\newcommand{\lineup}{\raisebox{5mm}[0mm][0mm]{\makebox[0mm][l]{\epsfxsize=9.0cm\epsfbox{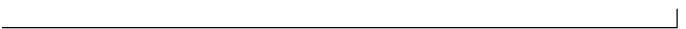}}}}
\newcommand{\linedown}{\raisebox{3mm}[0mm][0mm]{\makebox[0mm][l]{\hspace{9.0cm}\epsfxsize=9.0cm\epsfbox{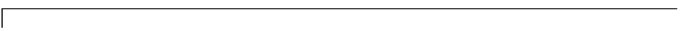}}}}
\newcommand{\half}{\frac{1}{2}}
\newcommand{\E}{\varepsilon}

\newcommand{\R}{\mbox{\rm I}\!\mbox{\rm R}}

\newcommand{\rme}{\mbox{e}}
\newcommand{\rmd}{\mbox{d}}

\newcommand{\be}{\begin{equation}}
\newcommand{\ee}{\end{equation}}
\newcommand{\bea}{\begin{eqnarray}}
\newcommand{\eea}{\end{eqnarray}}
\newcommand{\nn}{\nonumber}
\newcommand{\weiter}{\nonumber \\ & & }
\newcommand{\nicht}[1]{}
\newcommand{\eq}[1]{(\ref{#1})}

\newlength{\pcm}
\newlength{\pmm}
\newcommand {\GB} {\,\epsfxsize=1.2\pcm \parbox{1.2\pcm}{\epsfbox{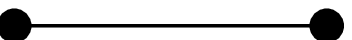}}\,}
\newcommand {\GH} {\,\epsfxsize=0.8\pcm \parbox{0.8\pcm}{\epsfbox{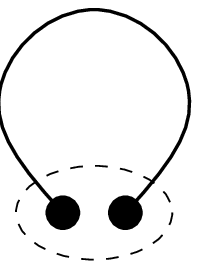}}\,}
\newcommand {\GM} {\,\epsfxsize=1.5\pcm \parbox{1.5\pcm}{\epsfbox{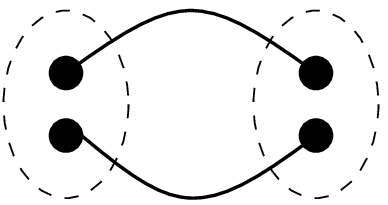}}\,}
\newcommand {\GO} {\,\epsfxsize=0.4\pcm \parbox{0.4\pcm}{\epsfbox{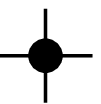}}\,}

\newcommand{\DynA}{{\,\epsfxsize=0.64\pcm \parbox{0.64\pcm}{\epsfbox{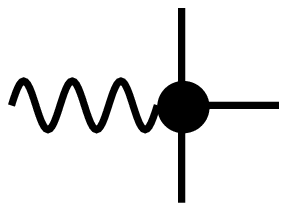}}\,}}
\newcommand{\DynB}{{\,\epsfxsize=1.28\pcm \parbox{1.28\pcm}{\epsfbox{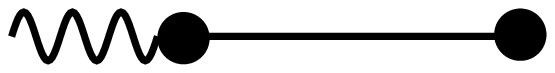}}\,}}
\newcommand{\DynC}{{\,\epsfxsize=1.04\pcm \parbox{1.04\pcm}{\epsfbox{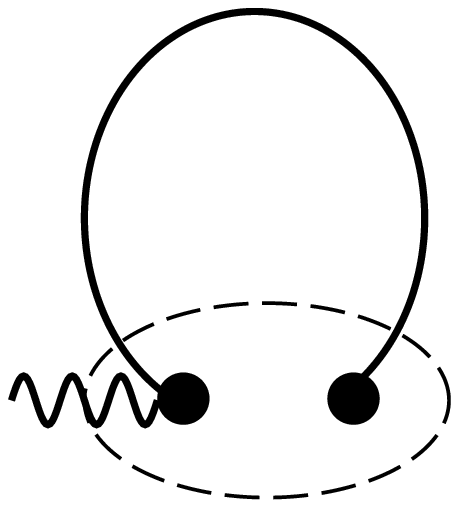}}\,}}
\newcommand{\DynD}{{\,\epsfxsize=1.44\pcm \parbox{1.44\pcm}{\epsfbox{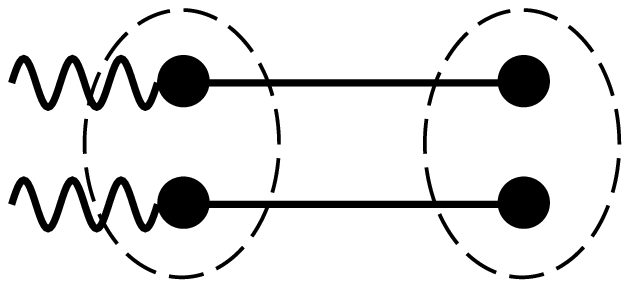}}\,}}
\newcommand{\DynE}{{\,\epsfxsize=0.64\pcm \parbox{0.64\pcm}{\epsfbox{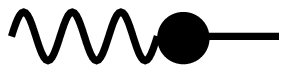}}\,}}
\newcommand{\DynF}{{\,\epsfxsize=0.8\pcm \parbox{0.8\pcm}{\epsfbox{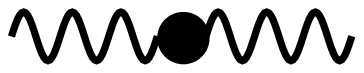}}\,}}

\title{Dynamics of Selfavoiding Tethered Membranes I \\
Model A Dynamics (Rouse Model)}
\author{\centerline{Kay J\"org Wiese}
\newline
\centerline{Fachbereich Physik, Universit\"at GH Essen,  45117 Essen,
Germany}}
\date{September 29, 1997}
\begin{document}
\maketitle
\begin{abstract}

The dynamical scaling properties of selfavoiding polymerized membranes
with internal dimension $D$ 
are studied using model A dynamics.
It is shown that the theory is renormalizable to all orders in 
perturbation theory and that the dynamical scaling exponent
$z$ is given by $z=2+D/\nu^*$. This result applies especially to
membranes ($D=2$) but also to polymers ($D=1$).
\end{abstract}

\begin{multicols}{2}
\noindent
Polymers and polymerized  flexible membranes show interesting statistical properties. Polymers have been investigated since long time, both static 
and dynamic \cite{CloJan90}.
Selfavoiding polymerized membranes have attracted a remarkable 
interest during the last years. Their static properties have 
been studied numerically 
\cite{AbrEtAl89,AbrNel90,GrestMurat90,GrestPetsche94,KrollGompper93}
experimentally
\cite{HwaKokufutaTanaka,SpectorEtAl94} and 
analytically \cite{AroLub87,KarNel87,DDG3,DavidWiese96a,WieseDavid96b}.
Dynamics  has  been regarded using scaling arguments
for polymers \cite{DeGennes75,DeGennes75b} and
membranes \cite{KantorKardarNelson1986b}.
For polymers, a renormalization group analysis has been performed
at 1- \cite{AlNoaimiEtAl78,OonoFreed81,Oono85,PuriSchaubOono86,WangFreed86,SchaubCreamerJo1988}
and 2-loop \cite{SchaeferPrivate} order.

For membranes, 
the analytical approach, inspired from polymer theory \cite{CloJan90}, 
relies on renormalization group and $\E$-expansion methods.
It was initiated in \cite{AroLub87,KarNel87}, where it was
used to perform calculations at 1-loop order.
Its consistency to all orders in perturbation theory has been established in \cite{DDG3,DDG4}.
Recently, 2-loop calculations have been performed, which give 
reliable results for all embedding dimensions \cite{DavidWiese96a,WieseDavid96b}.

In this letter we address the question of the dynamics of such membranes,
always including polymers as a special case.

The membrane is modeled by a continuum model \`a la Edwards:
the embedding of the $D$-dimensional membrane in $d$-dimensional bulk space
is described by the mapping $x\in \R^D\to  r(x)\in \R^d$.
The Hamiltonian, which describes the static
properties of the membrane, is
\begin{equation} \label{e:Ham}
\!\!\!\!{\cal H}[ r]= \frac{1}{2}\int_x\!\big(\nabla  r(x)\big)^2
+ b  \int_x\!\int_y\!
\delta^d\big ( r(x)- r(y)\big ) \ ,
\end{equation}
where $b$ is the coupling constant.
The dynamics of the membrane is given by the Langevin-equation
\be
\dot r(x,t)= - \lambda \frac{\delta H }{\delta r(x,t)} +
\zeta(x,t)
\ee
which models the purely diffusive motion of the membrane
(model A in the terminology of \cite{HoHa77}). 
The Gaussian noise $\zeta(x,t)$ has the correlation
\be 
 \overline{ \zeta(x,t) \zeta(x',t') } = 2 \lambda \delta^D(x-x')
\delta(t-t') \ .
\ee
This Langevin equation can be formulated using an effective field
theory \cite{Janssen92} with action
\be \label{e:J}
J\left[r,\tilde r \right]= \int_x \int_t \tilde r(x,t) \left( \dot r(x,t) +
\lambda \frac{\delta H }{\delta r(x,t)}   \right)
-\lambda \tilde r(x,t)^2 \ .
\ee
Expectation values are calculated by integrating 
over all 
fields $r$ and 
$\tilde r$. Prepoint-discretization is used and the average over
the noise has been taken. (This generated the term quadratic in 
$\tilde r$.) 

Perturbation theory is performed by expanding about the 
Gaussian theory. We use the free propagator $G$ and correlator $C$
in position-space
\bea
C(x,t)&:=&\frac1d \left< \half(r(x,t)-r(0,0))^2   \right>_0  \\
	&=&|x|^{2\nu}\, \Gamma^{-1}\left(-\nu \right) \times \nn\\
&& \ \!\times \left(-\frac1\nu \left({{4  \lambda |t|}\over{x^2}}
\right)^{\nu}+
\int_0^{{x^2}\over{4  \lambda |t|}} \frac{\rmd s}{s} s^{-\nu} 
\left( \rme^{-s}-1\right) \right) \nn
\\
G(x,t)&:=& \frac1d \left< r(x,t) \tilde r(0,0) \right >_0 \nn \\
	&=&
	\Theta(t) \,(4\pi \lambda |t|)^{-D/2}\,\rme^{-x^2/4\lambda |t|}\,
	S_D(2-D) \ ,
\eea
where $S_D$ is the volume of the unit sphere in $D$ dimensions,
\be
S_D=\frac{2 \pi^{D/2} }{\Gamma(D/2)}  
\ee
and 
\be
 \nu=\frac{2-D}2 \ .
\ee
Some normalization-factors have been absorbed into the measure in order 
to have 
\be
C(x,t) \approx |x|^{2-D} \quad \mbox{for} \quad x^2 \gg \lambda |t| \ . 
\ee
(For details compare appendix A of \cite{WieseDavid96b}.)
If  $x^2$ is much smaller than $\lambda |t|$, 
the correlator
approaches the finite value
\be
C(x,t) = \frac{(4 |t|\lambda)^{\nu}}{\Gamma(D /2)} + {\cal O}(x^2)
\ee
We should mention that the propagator is simply related to
the time-derivative of the correlator \be \label{e:PropCor}
G(x,t) = \Theta(t) \frac1\lambda \dot C(x,t)
\ee
We are now in a position to construct the perturbation-theory.
The interaction vertex is:
\be \label{e:DynBdef}
\DynB := 2 \int_k \tilde r(x,t) (ik)\,  \rme^{i k (r(x,t)-r(y,t))}
\ee
The perturbative expansion of an observable $\cal O$ can 
be written as
\be
\left< { \cal O} \right> = \mbox{Norm} 
\sum_n \frac {(\lambda b)^n} { n !} 
\int \left< {\cal O} \DynB^n \right>^c \ ,
\ee
where the normalization Norm has to be chosen so that $\left<1 \right>=1$
and the integral is taken over all arguments of the
interaction vertex. 
We claim that divergences only occur at short distances and short
times. To prove this look at a typical expectation value
\bea 
\left<{\cal O}\DynB^n \right>_0 &=& \sum_\alpha \int_{k_i} f_\alpha(x_l-x_m,t_l-t_m,k_l,k_m)
\times 
\nn\\
&& \quad \qquad \times\,\rme^{-\sum_{i,j} Q_{ij} k_i k_j  } \ ,
\label{14}
\eea
where each contribution consists of a function $f_\alpha$, which is a product 
of propagators, correlators and $k$'s and an exponential factor, with
\be
Q_{ij}=-C(x_i-x_j,t_i-t_j) \ .
\ee
$f_\alpha$ is a regular function of the distances. Divergences at 
finite distances can only
occur if   $Q_{ij}$ is not a positive form.
We will show that $Q_{ij}$ is a positive form for all $k_i$ which 
satisfy the constraint  
\be \label{neutral}
\sum_i k_i = 0 \ .
\ee 
This constraint always holds, see equation (\ref{e:DynBdef}). 
For equal times it is just the statement that the
Coulomb energy of a globally neutral assembly of charges is
positive. One simply identifies $C$ with the Coulomb-propagator and
$k_i$ with the charges.
In the dynamic case, write 
\bea
Q_{ij} &=&(2-D) S_D \int \frac{d^D\!p}{(2\pi)^D} \int\frac{\rmd \omega}{2\pi}\,\frac{2\lambda}{\omega^2 +\left(\lambda p^2\right)^2} \times
\weiter
\hspace{2cm}\times\left( \rme^{ip(x_i-x_j) + i\omega (t_i-t_j)} -1 \right)
\eea
\end{multicols}\widetext\noindent\lineup 
The exponential in \eq{14} now is
\begin{eqnarray} 
\sum_{i,j} k_i k_j Q_{ij} &=& (2-D) S_D  
\int \frac{d^D\!p}{(2\pi)^D}
\int \frac{\rmd \omega}{2\pi}\,
\frac{2\lambda}{\omega^2 +\left(\lambda p^2\right)^2}
\sum_{ij} k_i k_j 
\left( \rme^{ip(x_i-x_j)+i\omega (t_i-t_j)} -1 \right)
\nn\\
&=& (2-D) S_D  
\int \frac{d^D\!p}{(2\pi)^D}
\int \frac{\rmd \omega}{2\pi}\,
\frac{2\lambda}{\omega^2 +\left(\lambda p^2\right)^2}
\left|\sum_{i} k_i\, \rme^{ip x_i+\omega t_i}\right|^2
\end{eqnarray}
\begin{multicols}{2}\narrowtext
\noindent
\linedown
To get the second line, equation \eq{neutral} has been used. 
Note that again due to equation \eq{neutral}, 
the integral is ultraviolet convergent and thus 
 positive. It vanishes if and only if the charge-density, regarded as a function of space {\em and} time, vanishes. 
This is possible if and only if endpoints of the dipoles (which form
the interaction) are at the same point in space {\em and} time. 
No divergence occurs at finite distances. To renormalize the 
theory, only short distance divergences have to be 
removed by adding appropriate counter-terms. 
In addition, as the divergences occur at short distances,
they can be analyzed via a multilocal operator product expansion
(MOPE) \cite{DDG3}. For a detailed 
discussion of the MOPE and examples see \cite{DDG3,DDG4,%
WieseDavid96b,DavidWiese96a}.

We show now that the counter-terms which render the 
static theory \eq{e:Ham} finite are also sufficient for
the dynamical case \eq{e:J}. 
As an illustration we first calculate the 1-loop counter-terms for
the renormalization of the field and of the coupling-constant.
The first singular configuration appears, when both ends of the 
interaction vertex \eq{e:DynBdef} are contracted towards a single
point.  The leading term 
(MOPE-coefficient) of this expansion is (the dotted line
indicates points which are contracted):
\bea \label{e:1stdiag}
 \DynC 
&=& 2 \int_k -\left[ \tilde r(x,t)(ik)\right] \rme^{-k^2|x-y|^{2-D}}\times \nn\\
& &\qquad \times\left[(ik)(r(x,t)-r(y,t))\right] +\ldots
\eea
We now Taylor-expand $r(y,t)$ as 
\bea
r(y,t)&=&r(x,t)+(y-x)\nabla r(x,t) \nn\\
& &\  +\half \left[ (y-x)\nabla \right]^2
r(x,t) + {\cal O}(|x-y|^3)
\eea
The leading term in equation \eq{e:1stdiag} is
\bea
&&	\int_k \tilde r(x,t)  (-\Delta r(x,t)) \frac{(x-y)^2}{D} \frac{-k^2}{d} \rme^{-k^2|x-y|^{2\nu}} \nn\\
&&\quad = \tilde r(x,t) (-\Delta) r(x,t) \left( \frac{-1}{2D} \right)
|x-y|^{D-\nu d}
\eea
Denoting with $\bigg( \DynC \bigg| \DynA \bigg)$ the 
MOPE coefficient of $\DynC $ proportional to $\DynA = \tilde r(x,t) 
(-\Delta) r(x,t)$, this can be written in the form
\be
	\bigg( \DynC \bigg| \DynA \bigg) =  \bigg( \GH \bigg| \GO\bigg)
\ee
where 
\be
\bigg( \GH \bigg| \GO\bigg)=
-\frac1{2D} |x-y|^{D-\nu d}
\ee
is the static MOPE-coefficient \cite{DavidWiese96a,WieseDavid96b}.
This implies that the counter-term for the wave-function
renormalization is the same as in the static case. 
Let us now regard the counter-term for the coupling-constant
renormalization. Using the techniques of \cite{DDG3,DavidWiese96a,WieseDavid96b},
we obtain for the contraction of two interaction vertices to one
interaction vertex:
\bea
\bigg( \DynD \bigg| \DynB \bigg) &=& \frac d4 (G(x,t)+G(y,t)) 
\times \nn \\
&& \quad\  \times\left[ C(x,t)+C(y,t)\right]^{-d/2-1}
\eea 
$x$ and $y$ are the distances of the contracted endpoints 
of the dipoles, $t$ is their time-difference.
The trick is now to write this expression with the help of 
equation \eq{e:PropCor} as 
\be
- \frac1{2 \lambda} \Theta(t>0) \frac \rmd {\rmd t} \left[ C(x,t) 
+C(y,t) \right]^{-d/2} 
\ee
In performing the perturbation-theory, we have to integrate over
all times. If we use, as is usually done, no cut-off in the time-direction,
the time integral will simply give the value of the function 
at its lower bound:\be
\int_0^\infty \rmd t \, \bigg( \DynD \bigg| \DynB \bigg) 
= \frac1{2\lambda}\bigg( \GM \bigg| \GB \bigg)
\ee 
(The r.h.s.\ is the counter-term of the static theory,
see \cite{DDG3,DavidWiese96a,WieseDavid96b}.)
We easily convince ourselves that this relation implies the same
counterterm as in the static case, if we take care of the 
additional combinatorial factor 2 for the time-ordering of
the interaction vertices. 

One knows from general arguments that the divergences associated to 
short distances in space are removed by the static counter-terms
\cite{Zinn}.
We now use the fact that the static theory is renormalizable
\cite{DDG3}. This implies that new divergences can only
appear for short times. We therefore have to analyse all
possible divergences of this type. Using the MOPE, the most 
relevant divergences are associated to the operator
\be \label{DynE}
 \DynE= \tilde r(x,t) \dot r(x,t) \ . 
\ee
We now regard a general contraction of $n$ dipoles towards
$\DynE$:
\be
\DynB^n \ \longrightarrow \DynE 
\ee
In order to obtain the operator $\DynE$, one has to contract
all fields $r$ and $\tilde r$ except of the field $\tilde r(z,t)$ with
the largest time-argument (all other contractions give 0). One also
has to leave uncontracted one arbitrarily chosen
field $r$. Due to the structure of the interaction
\eq{e:DynBdef}, the field $r$ always appears in the form
$r(x,t-\tau)  - r(y,t-\tau)$.
So the contraction yields
\be
\tilde r(z,t) \left[r(x,t-\tau)-r(y,t-\tau)\right] M(\mbox{distances}) \ ,
\ee 
where $M$ denotes the MOPE-coefficient which depends on the distances
in space and time.
Now, $r(x,t-\tau)-r(y,t-\tau)$ has to be expanded about $(z,t)$.
The leading term has at least {\em one} spatial gradient. 
No term of the form of \eq{DynE} can be constructed. 
Therefore, there is no singular contribution of this type in any order in 
perturbation theory and no renormalization of $\DynE$ is needed.
The last at $\E=0$ marginal operator is 
\be
	\DynF = \tilde r(x,t)^2 \ .
\ee
Its renormalization is given by the  
fluctuation-dissipation theorem, see below.

We now introduce  renormalized quantities according to
\bea
r &=&\sqrt{Z} r_R \nn\\
\tilde r &=& \sqrt{\tilde Z} \tilde r_R\nn\\
\lambda &=& Z_{\lambda} \lambda_R€\nn\\
b&=& b_R Z_bZ^{d/2}\mu^\E 
\eea
$\E$ is the dimensional regularization parameter, defined by
\be
\E= 2D- \nu d \ .
\ee
The fluctuation-dissipation theorem states that
\be
\Theta(t) \frac\partial {\partial t}\bigg< \half  \left( r(x,t)-r(0,0)\right)^2\bigg>=\lambda \bigg<  r(x,t)\tilde r(0,0) \bigg>
\ee
This relation for the full expectation-values holds  as well for renormalized
as for bare quantities. We therefrom deduce that
\be
Z_{\lambda} = \sqrt{\frac{Z}{\tilde Z}} 
\ee
In addition, our findings that the term $\DynE$ has not to be renormalized
indicate that 
\be
\sqrt{Z \tilde Z}=1 \ .
\ee 
It follows that
\be 
Z_{\lambda}=Z \ .
\ee
As usual,
the renormalization group $\beta$-function $\beta(b_R)$ and the full scaling
dimension $\nu(b_R)$ of $ r$ are obtained from the  variation of the 
coupling constant 
and the field with respect to the renormalization scale $\mu$,
keeping the bare couplings fixed. They are written in terms of
$Z$ and $Z_b$ as 
\begin{eqnarray}
\label{e:beta}
\beta(b_R) &=&
\frac{-\E b_R} {1+ b_R\frac{\partial}{\partial b_R } \ln Z_b +
\frac{d}2 b_R \frac{\partial}{\partial b_R} \ln Z}\\
\label{e:nu}
\nu (b_R) &=&
\frac{2-D}2 -\half \beta(b_R) \frac{\partial}{\partial b_R} \ln Z 
\end{eqnarray}
The $\beta$-function has an IR-fixed point for $b_R^*>0$. The function
$\nu(b_R)$ is for large scales therefore given by
\be
\nu^*=\nu(b^*) \ .
\ee
We deduce that the correlation-function scales for equal times as
\be
\left< (r(x,0)-r(0,0))^2 \right> \sim |x|^{2\nu^*} \ .
\ee
If we define the exponent $z$ for the auto-correlation-function as
\be
\left< (r(0,t)-r(0,0))^2 \right> \sim |t|^{2/z} \ ,
\ee
the exponents  $z$ and $\nu^*$ are dependent:
\be \label{theresult}
z=2+\frac{D}{\nu^*}
\ee
For polymers (D=1), this relation was given 
using scaling arguments in  \cite{DeGennes75}, for membranes ($D=2$)
in \cite{KantorKardarNelson1986b}. 
This result was followed by perturbative 
calculations for polymers in 1-loop \cite{AlNoaimiEtAl78,OonoFreed81,Oono85,PuriSchaubOono86,WangFreed86,SchaubCreamerJo1988} and
2-loop \cite{SchaeferPrivate} order. 

It is interesting to note that for polymers and membranes 
\eq{theresult} can be written in the form
\be \label{theresult2}
z=2+d_f
\ee
where $d_f$ is the fractal dimension of the membrane or the polymer.

It seems attractive to use \eq{theresult2} in Monte Carlo 
simulations to improve the determination of $d_f$ for
selfavoiding membranes. 
This might be advantagous as
$z$ can be measured for any point and
any time, i.e. on a large statistical ensemble. 

In conclusion: We have shown that the purely diffusive motion 
 of polymers and polymerized membranes is given to all orders
in perturbation theory by \eq{theresult2}. 
Experimentally however, an additional hydrodynamic interaction
is present. This question will be addressed in a subsequent 
publication.

\acknowledgments
\noindent 
It is a pleasure to thank F. David, H.~W. Diehl and L. Sch\"afer 
for stimulating discussions. 

\bibliography{../citation/citation}

\bibliographystyle{../macros/KAY}

\end{multicols}
\end{document}